# DAΦNE INJECTION SYSTEM UPGRADE

A. Ghigo, LNF-INFN, Frascati, Italy

*Abstract*

High luminosity in DAΦNE [1] needs very high electron and positron currents stored. A full energy (510 MeV) injection system composed by a full energy electron and positron linac and an accumulator-damping ring is presently used. The electron and positron beams, alternatively accelerated by the linac, are injected and stacked in the accumulator with high efficiency thanks to its large acceptance and short damping time. The damped beams are extracted and transferred to the main ring through a long transfer line that has been built inside already existing buildings. The refill time of the collider is limited by the transfer line set-up change between the two different beams modes. In this paper a transfer line modification is proposed in order to reduce the switch time. A possible injection scheme for the main rings is also described.

## 1 INTRODUCTION

The DAΦNE High Luminosity option foresees a very powerful injection system to allow injection into the collider of very high average currents, distributed in 150 bunches, in very short time. To obtain high average luminosity, high efficiency injection is mandatory because the beam lifetime is limited to few minutes, due both to the strong contribution of the Touschek effect enhanced by the very small longitudinal and transverse bunch dimensions and to the beam-beam bremsstrahlung [1].

An injection system composed by a full energy electron and positron linac and an accumulator-damping ring is presently used [2,3]. This system provides injection of 1A of electron current in less than 1 min, distributed in 120 bunches; injection of 1A of positrons needs approximately the same time.

The electron and positron beams are transferred to the main ring through one long transfer line passing through existing building. At least 3 minutes are necessary for switching the transfer line from electron to positron mode, and vicecersa, limiting the refill speed.

Two separate lines to transfer the beams from the accumulator to the main rings are here proposed. The injection time will be decreased and the additional possibility to alternatively inject corresponding electron and positron bunches will be gained.

## 2 PRESENT INJECTION SCHEME

*High repetition rate injection*

Both KLOE and FINUDA experiments on DAΦNE run in top-up mode, with a refill every 15-20 min. Data taking is disabled in the KLOE detector for only 50 ms (3 damping times) at each injection shot, by forbidding the acquisition trigger. Therefore the acquisition duty cycle is close to 100%. During the switch of the injection system from electron to positron mode and viceversa there is a significant reduction of the stored currents of both beams since the lifetime is of the order of half an hour. The peak currents of the two beams are not reached at the same time and the average luminosity decreases not only due to the total current reduction but also to the current difference between corresponding bunches.

The injection system switches from electron to positron mode with the following operations:

- The tungsten target in which the electron-positron pairs are produced is inserted in the Linac, intersecting the electron trajectory.
- The gun current is increased up to 6 A and a quadrupole triplet focusing the beam on the target is switched on. The capture section solenoids placed after the target are powered.
- The accelerating sections phases are shifted to positron accelerating mode.
- The positron beam produced and accelerated in the Linac, uses the same transfer line for injection into the Accumulator ring and (partially) to pass from the Accumulator to the DAΦNE main rings. The magnetic field of the bending magnets, placed in the common transfer line, must change polarity in going from electron to positron mode.
- The set up of all quadrupole magnets is changed.

The most time consuming operation is the polarity switch and set up of the transfer line magnets between electron and positron injection.

## 3 UPGRADED INJECTION SYSTEM

The proposal of two new transfer lines, one for the electron beam and one for positron one, between the Damping Ring and the Main Rings avoid the magnets polarity switch passing from electron to positron mode.

The only common line to electrons and positrons will be from the linac to the damping ring; there will be still one pulsed dipole magnet in it. The time between two polarity switches is of the order of 100 ms, equivalent to 5 damping times in the damping ring. This common line will have constant quadrupoles currents settings, which match the betatron planes at the damping ring input in the two short separate arms placed before the septum magnets. The Damping Ring has a symmetric configuration and it is used injecting the electron beam from one side through CW septa and extracting the damped beam through the septa dedicated to positron injection; the same configuration is used for positrons the other way round. The ring has a compact design, its 32 m of circumference are very crowded and modifications should be avoided.

In order to deviate the beam extracted from the DR to the new lines that transfer the beams to the main rings two additional pulsed magnets are required.

The current pulse duration in these magnets is 100 ms, the same as in the linac to damping ring pulsed magnet. The kick angle should allow extraction of the beams from the present lines to the new separated ones: it is of the order of 6 degrees, as presently delivered by the existing pulsed magnet used in the Linac to deflect the beam towards the energy analyser.

The two lines exit from the DR hall passing above the ring and partially over the existing transfer line. As shown in Fig. 1, a new short shielded tunnel must be realized in order to minimize the path, since there is not enough space in the concrete tunnel between the linac tunnel and the damping ring hall to host two new lines.

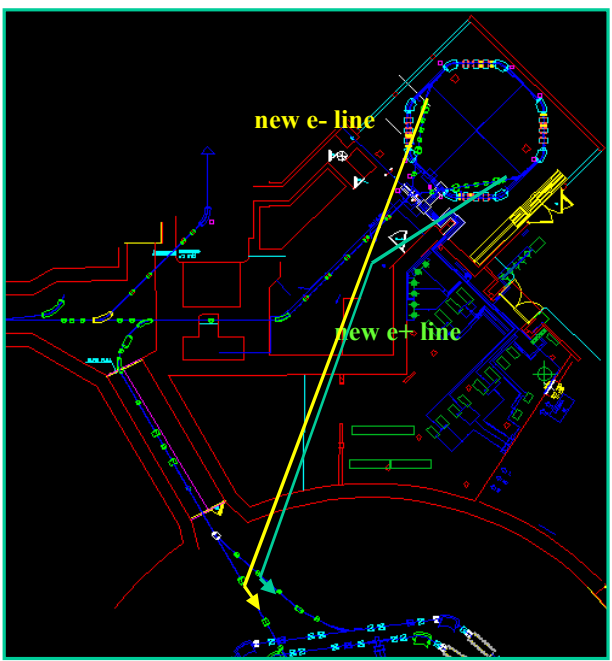

Figure 1: New transfer lines layout from the damping ring to the present DAΦNE Main Rings.

The foreseen location for the new tunnel has just been classified as controlled area.

The straight portion of these transfer lines could be in common for electrons and positrons with two Y magnets at the ends merging and separating the beams into the main rings injection lines. With this configuration the injection switch sequences will be:

- The positron target is kept inserted also for electrons, and the capture section solenoids are always powered.
- Switch from positron to electron mode consists therefore only in shifting the accelerator sections phases.
- The pulsed magnet between the linac and the damping ring switches polarity in less than 100 ms.
- The damping ring injection kickers change the pulser charging voltage in less than 20 ms.

The extraction sequence will be:

- The damping ring kicker magnets change the charging voltage.
- The pulsed magnet downstream the extraction septum exit is switched on, sending the extracted beam to the damping ring – main ring transfer line.
- The bunch is transferred to the corresponding main ring.

With this hardware configuration and operation procedure the dead time in the injection switch is avoided. The injection of interacting electron and positron bunches is also possible only by switching the pulsed magnets and the linac phases at each shot. This injection scheme allows to have equal currents in the interacting bunches, thus optimising instantaneous and average luminosity.

This upgrade of the injection system could be very useful also for the present DAΦNE configuration while it is mandatory for the high luminosity option.

## 4 DAΦNE II MAIN RINGS INJECTION

The high luminosity upgrade of DAΦNE is based on the strong RF focusing principle [4], in which the bunch length changes along the ring passing from the maximum length in the high voltage RF cavities section to a minimum in the interaction region.

The vacuum components that contribute to the beam coupling impedance should be concentrated in the section where the bunch length is maximum. In the preliminary ring design a long straight section for each ring is dedicated to the RF cavities and the injection section.

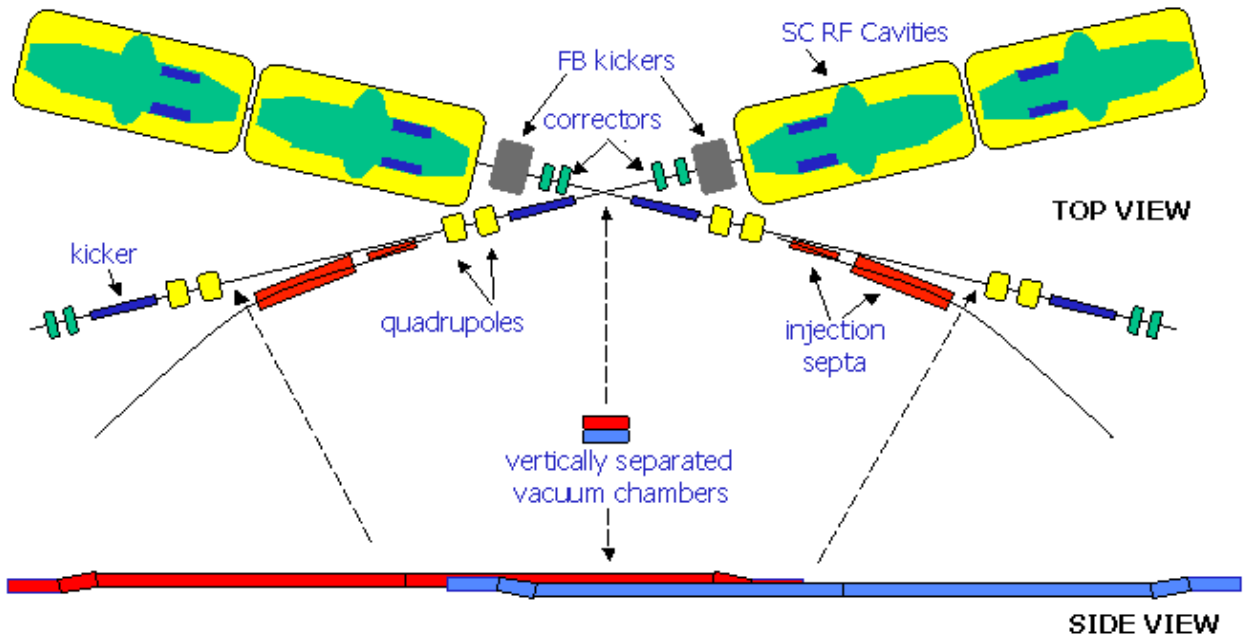

Figure 2: Crossing section schematic layout.

The electron and positron beams are injected into the Main rings by means of CW septa and pulsed kicker magnets.

In order to keep the beams well separated in the area in which the two rings cross, two vertically separated vacuum chambers are proposed [see Fig. 2].

Vertical dipoles inside the injection straight sections are used to separate the vacuum chambers of the electron and positron rings at the crossing point (see Fig. 2). The vertical separation must be the smallest possible in order to minimize the induced vertical dispersion.

## 5 CONCLUSIONS

A modification of the injection transfer lines is proposed in order to avoid the dead time due to the switch between electron and positron modes in the present configuration.

Already in the present DAΦNE configuration the integrated luminosity would benefit from such an upgrade. For the DAΦNE high luminosity option the new transfer lines are mandatory to allow continuous injection.